\documentstyle[12pt]{article}

\begin{document}

\setcounter{page}{1}

\pagestyle{plain} \vspace{1cm}
\begin{center}
\Large{\bf A Revision to the Issue of Frames by Non-minimal Large Field Inflation }\\
\small \vspace{1cm} {\bf M. Shokri \footnote{mehdi.shokri@uniroma1.it}}\\
\vspace{0.5cm} {\it Physics Department and INFN,\\
Universit`a di Roma La Sapienza, Ple. Aldo Moro 2, 00185, Rome, Italy}
\end{center}\vspace{1.2cm}
\begin{abstract}
We present an extended study of inflationary models that inflaton field is non-minimally coupled with gravity. We study parameters space of the models up to the second (and in some cases third) order of the slow-roll parameters for usual large field potentials in Jordan and Einstein frames that are connected each other by conformal transformation. We calculate inflationary parameters and the results are compared in both frames and also with observations. By using the recent observational datasets, we present a discussion in order to clarify the physical frame between Jordan and Einstein frames. Also, some suggestions are expressed in order to navigate us for the future works.
\end{abstract}
\newpage
\section{Introduction}
Inflation as accelerating phase in the early universe can dissolve the defects of the big bang cosmology such as flatness, horizon and monopole problems. Moreover, it provides a good mechanism to generate the primordial density perturbations for structure formation of the
universe and also tensor perturbations which are responsible to produce the primordial gravitational waves [1-3]. Despite these successes, the idea of inflation suffers from several ambiguities which the most important of them is the initial condition. We believe that inflationary epoch has emerged from Planck era, but we have no enough knowledge about how it is happened. Another important problem is that there is no unique model to explain inflation. To now, many versions of inflation have been proposed: single scalar field inflation, multi field inflation (in particular hybrid inflation), modified gravity models of inflation, non-minimal inflation and etc. Nevertheless, in this work we focus on the inflationary models with non-minimal coupling (NMC). The inflationary cosmology believes that a single scalar field (\textit{inflaton}) is responsible to drive inflation and there is no any kind of matter in that era. When we look at the form of action in General Relativity (GR) for inflation, there is a minimal coupling (MC) between Ricci scalar as gravitational sector and scalar field as matter sector. According to the NMC idea, we can not disregard direct interaction between these two quantum fields [4]. Therefore, the form of action will be modified by adding the NMC term $\frac{1}{2}\xi R\varphi^{2}$ as
\begin{equation}
S=\int{d^{4}x\sqrt{-g}\bigg(\frac{R}{2\kappa^{2}}-\frac{1}{2}g^{\mu\nu}\partial_{\mu}\varphi\partial_{\nu}\varphi+V(\varphi)+\frac{1}{2}\xi
R\varphi^{2}\bigg)}
\end{equation}
where $V(\varphi)$ is potential of scalar field and $\xi$ depicts the coupling constant that its value is profoundly effective on the viability of an inflationary scenario. Because of some difficulties, using the NMC idea is not delicious for us. But it is forced upon our cosmological situations. In principle, NMC arises in the presence of the quantum corrections of scalar field, even it is necessary at the classical level for renormalization of scalar field theory in curved space [5]. As a result of this reality, the NMC term is unavoidable in many cosmological investigations. It is pointed out that the NMC term can be included other forms of $\varphi$ not only $\varphi^{2}$. Nevertheless, in this paper we use the ready shape because our purpose is concentrated on the conformal transformation and the issue of frames. Although this form is beneficial for us, sometimes it can be distractive. For example, some authors consider a pure mass term $\frac{1}{2}m^{2}\varphi^{2}$ for inflaton in the action (1) that its shape is distorted by the NMC term. Therefore, it is more difficult to achieve the slow-rolling scalar field. In fact, the NMC term plays the role of an effective mass for inflaton. About the value of coupling constant $\xi$, there are two approaches, generally. The common attitude in the literature on non-minimally coupled scalar fields is that $\xi$ is a free parameter to fine-tune at one's own will in order to solve problems of the inflationary scenario under consideration [6]. Sometimes this standpoint is not suitable for our physical situations because of its problems. Therefore, we can focus on a more rigorous approach. It consists the range of values of $\xi$ that are fixed by particles physics [7]. In fact, the first approach is regarded when we have no prescription for $\xi$. The value of $\xi$ in this attitude depends on the nature of inflaton and on the theory of gravity under consideration.\\
In the presence of the NMC, we are faced with many difficulties in our inflationary models. Therefore, we can use the conformal transformation
\begin{equation}
\hat{g_{\mu\nu}}=\Omega g_{\mu\nu}
\end{equation}
to escape them due to the fact that it is often used as a mathematical tool to map the equations of motion of physical systems into mathematically equivalent sets of equations that are more easily solved and computationally more convenient to study. In the equation (2), conformal factor $\Omega=\Omega(\varphi(x))$ is a differentiable and non-zero function. The conformal transformation is involved with many gravitational theories that are based on scalar field because of the dependence of $\Omega$ to scalar field. Theories include scalar-tensor and non-linear theories of gravity, Kaluza-Klein theories, fundamental scalar fields include Pseudo-Nambu-Goldstone bosons (PNGB), Higgs bosons and dilatons in supergravity theories and many inflationary scenarios that are built on scalar field [8-13]. In the present work, we use the conformal transformation (2) for the inflationary models with NMC. Unfortunately, using the conformal transformation is haunted by confusion and ambiguities in particular the issue of frames that this work is dedicated it. By the conformal transformation, we transfer the parameters space from \textit{Jordan frame} (the main frame) to \textit{Einstein frame} (the conformal frame) that is easier to study. We expect that the results of two frames be the same (physically equivalence) because they are mathematically equivalent. At the classical level, two frames are somewhat equivalent, but it seems that in the high energy level and in the presence of quantum corrections, the physical equivalence between two frames will be broken. Now, we are encountered with important questions. As example, why the conformal symmetry (the physical equivalence) breaks when the quantum effects are considered? and if we accept the discrepancy between the results of two frames, which of them is the physical frame?.\\
The subject is still open and there are some works that are specified to the frames problem [14-23]. Our purpose in this paper is the presentation of a better judgment about the physical frame between Jordan and Einstein frames by non-minimal inflation. All above information give us motivation to arrange the paper as follow. In section 2, we describe inflationary analysis for two potentials in two frames, separately. The obtained results compare with observations in section 3. In the section 4, we present a quantitative discussion and outlook about the issue of frames based on the obtained result. Eventually, the section 5 is dedicated to the summary.
\section{The Inflationary Analysis in Two Frames}
In this section, we present inflationary analysis for two large filed potentials in Jordan and
Einstein frames, respectively.
\subsection{Jordan Frame}
The form of action in this frame (1) includes the explicit term of NMC between the Ricci scalar
and scalar field. By varying it with respect to $\varphi$, the Klein-Gordon equation is found as
\begin{eqnarray}
\ddot{\varphi}+3H\dot{\varphi}+\bigg(\frac{\kappa^{2}\xi\varphi^{2}(1+6\xi)}
{1+\kappa^{2}\xi\varphi^{2}(1+6\xi)}\bigg)\frac{\dot{\varphi}^{2}}{\varphi}=\bigg(\frac{4\kappa^{2}\xi\varphi
V(\varphi)-(1+\kappa^{2}\xi\varphi^{2})\frac{dV}{d\varphi}}{1+\kappa^{2}\xi\varphi^{2}({1}+6\xi)}\bigg)
\end{eqnarray}
where overdots denote time derivatives. Also, the variation with
respect to the metric leads to the Einstein equation as
\begin{equation}
(1+\kappa^{2}\xi\varphi^{2})G_{\mu\nu}=\kappa^{2}\tilde{T}_{\mu\nu}
\end{equation}
where
\begin{equation}
\tilde{T}_{\mu\nu}=\nabla_{\mu}\varphi\nabla_{\nu}\varphi-\frac{1}{2}g_{\mu\nu}\nabla^{\gamma}\varphi\nabla_{\gamma}\varphi+Vg_{\mu\nu}-\xi\bigg(g_{\mu\nu}\Box
(\varphi^{2})-\nabla_{\mu}\nabla_{\nu}(\varphi^{2})\bigg).
\end{equation}
It is obvious that for $\xi=0$, we obtain the familiar equations in GR. The equation (4) can be
rewritten in two different approaches. In the first, we deal with an effective and $\varphi$-dependent
gravitational constant as
\begin{equation}
G_{eff}\equiv\frac{G}{(1+\kappa^{2}\xi\varphi^{2})}
\end{equation}
which this attitude is analogous to the Brans-Dicke theory because
scalar field $\varphi_{BD}$ is equal to the inverse of an
effective gravitational constant
$G_{\varphi}=\frac{1}{\varphi_{BD}}$. In this approach, the
Einstein equation (4) is rewritten as
\begin{equation}
G_{\mu\nu}=\kappa_{eff}^{2}\tilde{T}_{\mu\nu}
\end{equation}
where $\kappa_{eff}^{2}\equiv8\pi G_{eff}$. In the second
viewpoint, we are faced with a $\varphi$-independent gravitational
constant $G$, so the equation (4) is found as
\begin{equation}
G_{\mu\nu}=\kappa^{2}{T}_{\mu\nu}
\end{equation}
where
\begin{equation}
T_{\mu\nu}\equiv\frac{\tilde{T}_{\mu\nu}}{(1+\kappa^{2}\xi\varphi^{2})}.
\end{equation}
It is pointed out that employing the approaches (7) and (8) introduces two critical values of scalar field for $\xi<0$ as
\begin{equation}
\pm\varphi_{critical}=\pm\frac{1}{\kappa\sqrt{|\xi|}}
\end{equation}
which are barriers that scalar field can not cross them. In fact, the transition from the equation (4) to eqs. (7) or (8) leads to encounter with a restricted class of solutions. Another important point is that the conservation equation for the energy-momentum tensor is valid for the equation (8) due to the contracted Bianchi identities $\nabla^{\nu}G_{\mu\nu}=0$. But, for the origin form of the Einstein equation (4), it is preserved only for the case of $\varphi=const$.\\
To achieve the Friedmann equations in the presence of NMC, we assume a spatially flat FRW cosmology with line element $ds^{2}=g_{\mu\nu}dx^{\mu}dx^{\nu}=dt^{2}-a^{2}(t)d\vec{x}^{2}$. Therefore, the equations take the following form
\begin{equation}
H^{2}=\frac{\kappa^{2}}{3(1+\kappa^{2}\xi\varphi^{2})}\bigg[\frac{\dot{\varphi}^{2}}{2}+V(\varphi)-6\xi
H\varphi\dot{\varphi}\bigg].
\end{equation}
The majority of conventional inflationary scenarios are involved with the slow-roll approximation. According to this approximation, inflaton slow rolls from the start point of inflation to the end point of it so that its conditions in the Hamilton-Jacobi formalism are defined as
\begin{equation}
\bigg|\frac{\ddot{\varphi}}{\dot{\varphi}}\bigg|\ll H,\quad\quad
\bigg|\frac{\dot{\varphi}}{\varphi}\bigg|\ll H,\quad\quad
\dot{\varphi}^{2}\ll V(\varphi),\quad\quad \bigg|\dot{H}\bigg|\ll
H^{2}.
\end{equation}
Note that in this frame, we deal with the Hamilton-Jacobi formalism of the slow-roll approximation. Also, the slow-roll parameters play a vigorous role in our setup and we can find them as
\begin{equation}
\epsilon\equiv\frac{-\dot{H}}{H^{2}},\quad\quad
\eta\equiv\frac{-\ddot{H}}{H\dot{H}},\quad\quad
\zeta\equiv\frac{V'\delta\phi}{\dot{\phi}^{2}}=\frac{V'H}{2\pi\dot{\phi}^{2}}
\end{equation}
where overdots depict time derivatives and prim implies to $\frac{d}{d\varphi}$. It should be kept in the mind that during inflationary phase, the slow-roll parameters remain less than unity and inflation ends when the condition $\xi=1$ or $\eta=1$ is fulfilled. In order to keep the analysis, we focus on two usual inflationary potentials chaotic inflation and power-law potentials, respectively.
\subsubsection{\textbf{Chaotic Inflation Potential}}
A pure mass term
\begin{equation}
V(\varphi)=\frac{1}{2}M^{2}\varphi^{2}
\end{equation}
is the simplest potential for inflaton and also is an appropriate
example to exhibit the manner of even potentials. In the form of this potential, $M\sim10^{-6}m_{pl}$ is the mass of inflaton.
In fact, in the presence of NMC term, we deal with an
effective mass $m_{eff}=(m^{2}+\xi R)^{\frac{1}{2}}$. Note that
there is no inflationary solution for the large positive value of
$\xi$, so we restrict ourselves to $\xi\leq\frac{1}{6}$. The field equations (3) and (11) for this potential in presence of
the slow-roll conditions (12) are found as
\begin{equation}
H^{2}\approx\frac{\kappa^{2}M^{2}\varphi^{2}}{6(1+\kappa^{2}\xi\varphi^{2})},\,\,\
3H\dot{\varphi}\approx\frac{M^{2}\varphi(\kappa^{2}\xi\varphi^{2}-1)}{1+\kappa^{2}\xi\varphi^{2}({1}+6\xi)}.
\end{equation}
By using above equations, the slow-roll parameters (13) are
calculated as
\begin{equation}
\epsilon=\frac{2({1}-\kappa^{2}\xi\varphi^{2})}{\kappa^{2}\varphi^2({1}+\kappa^{2}\xi\varphi^{2}({1}+6\xi))},
\end{equation}
\begin{equation}
\eta=-\frac{8\xi}{({1}+\kappa^{2}\xi\varphi^{2}({1}+6\xi))}+\frac{4\xi(1+6\xi)(\kappa^{4}\xi^{2}\varphi^{4}-1)}{({1}+\kappa^{2}\xi\varphi^{2}({1}+6\xi))^2},
\end{equation}
and
\begin{equation}
\zeta=\frac{9M\varphi^{2}}{{2}{\pi}}(\frac{\kappa^{2}}{{6}(1+\kappa^{2}\xi\varphi^{2})})^{3/2}
\bigg[\frac{({1}+\kappa^{2}\xi\varphi^{2}({1}+6\xi))}{(\kappa^{2}\xi\varphi^{2}-1)}\bigg]^{2}
\end{equation}
where $\kappa^{2}=8\pi G$. For our purpose, the obtained slow-roll parameters should be depended only
to $\xi$. Therefore, we consider these parameters for the beginning and the end of inflation by
definition of new quantities $\beta$ (for the end of inflation) and $m$ (for the time of Hubble Crossing
(HC)) as
\begin{equation}
\beta^{2}(\xi)=\kappa^{2}\xi\varphi^{2}_{e}
\end{equation}
and
\begin{equation}
m^{2}(\xi)=\kappa^{2}\xi\varphi^{2}_{HC}
\end{equation}
where $\varphi_{HC}$ and $\varphi{e}$ imply to the value of
infalton at the time of HC and the end of inflation, respectively.
We know that inflationary period ends when $\epsilon=1$ or $\eta=1$ be satisfied. Hence, by combination of the equation (19) and the rewritten slow-roll parameters at the end of inflation, we arrive to the explicit expression for $\beta$. But, the situation for the time of HC is very different because the initial condition of inflation is not clear, completely. Therefore, we assume three
different confined cases for the time of HC. They include the \textit{Exact Case} which there is no any
comparison between the value of inflaton at time of HC and the end of inflation, the \emph{Case of $m\gg\beta$} which the value of inflaton at the beginning of inflation is much bigger than its value when inflation ends and the \emph{Case of $m\ll\beta$} which the value of inflaton at the start of inflation is much smaller than its value when inflation ends. By setting the condition of $\epsilon=1$ in the equation (16) and by using the equation (19), $\beta$ takes the following form
\begin{equation}
\beta^{2}=\frac{-(1+2\xi)}{2(1+6\xi)}\pm\frac{\sqrt{1+12\xi+52\xi^{2}}}{2(1+6\xi)}
\end{equation}
and also we can rewrite the slow-roll parameters at the time of HC
by using the equation (20) as
\begin{equation}
\epsilon=\frac{2\xi({1}-m^{2})}{m^2({1}+m^{2}({1}+6\xi))}
\end{equation}
\begin{equation}
\eta=-\frac{8\xi}{({1}+m^{2}({1}+6\xi))}+\frac{4\xi(1+6\xi)(m^{4}-1)}{({1}+m^{2}({1}+6\xi))^2},
\end{equation}
and
\begin{equation}
\zeta=\frac{9Mm^{2}}{{2}{\pi}\kappa^{2}\xi}(\frac{\kappa^{2}}{{6}(1+m^{2})})^{3/2}(\frac{{1}+m^{2}({1}+6\xi)}{m^{2}-1})^{2}.
\end{equation}
The number of e-folds has great role in inflationary analysis and it can be defined as
\begin{equation}
N=\int^{t_{f}}_{t_{i}}Hdt=\int^{\varphi_{e}}_{\varphi_{i}}\frac{H}{\dot{\varphi}}d\varphi
\end{equation}
where subscripts $e$ and $i$ denote to the end and the start of inflation, respectively. The fact is
that the scales of interest to us crossed outside of the horizon approximately 60-e-folds before
the end of inflation. The number of e-folds for this potential in the Jordan frame takes the
following form as
\begin{equation}
N=\frac{1}{8\xi}\ln\bigg({\frac{(\beta^{2}-1)(m^{2}+1)}{(\beta^{2}+1)(m^{2}-1)}}\bigg)+\frac{(1+6\xi)}{8\xi}\ln\bigg({\frac{\beta^{4}-1}{m^{4}-1}}\bigg)
\end{equation}
since the above equation has no analytical solution to $m$, by expanding it and keeping the first order terms, $N$ is found as
\begin{equation}
N\approx\frac{1}{4\xi}(m^{2}-\beta^{2})+\frac{(1+6\xi)}{8\xi}(m^{4}-\beta^{4}).
\end{equation}
As we noticed before, the parameters (22) to (24) can be rewritten for the three different types
of the initial condition, separately. Now, we focus our attention to those and also find some
phrases for $m$.
\begin{center}
$\bullet\textit{\textbf{Exact Case}}$
\end{center}
In this case, there is no comparison between the values of inflaton in two points (the start and
the end of inflation), so we deal with the perfect form of the equation (27) and for $m$, we find
\begin{equation}
m^{2}=-\frac{1}{(1+6\xi)}\pm
\frac{1}{(1+6\xi)}\sqrt{1+2\beta^{2}+12\beta^{2}\xi+8\xi
N+48\xi^{2}N+\beta^{4}+12\beta^{4}\xi+36\beta^{4}\xi^{2}}.
\end{equation}
We note that suitable choice for $m^{2}$ depends to the value and the sign of $\xi$ and also to the
positive definition of $m^{2}$ and $\beta^{2}$. The form of the first order of spectral index in the Jordan
frame is as
\begin{equation}
n_{s}=1-6\epsilon+2\eta
\end{equation}
and by combination of eqs. (22), (23) and (28), we have
\begin{eqnarray}
n_{s}=1-\frac{(12\xi(\sqrt{\Psi}-2)-72\xi^{2})}{\sqrt{\Psi}(1-\sqrt{\Psi})}-\frac{16\xi}{\sqrt{\Psi}}
+\frac{8\xi\bigg((-1+\sqrt{\Psi})^{2}-(1+6\xi)^{2}\bigg)}{(1+6\xi)\Psi}.
\end{eqnarray}
The form of the second order of spectral index is given as
\begin{eqnarray}
n_{s}=1-6\epsilon+2\eta+\frac{1}{3}(44-18c)\epsilon^{2}
+(4c-14)\epsilon\eta+\frac{2}{3}\eta^{2}+\frac{1}{6}(13-3c)\zeta^{2}
\end{eqnarray}
where $C\equiv4(ln2+\gamma)$ with $\gamma=0.577$. By putting the eqs. (22) to (24) and (28), we can reach
to the form of the second order of spectral index in this limited case. Another inflationary
parameter is the first order of running of spectral index
\begin{equation}
\alpha_{s}=16\epsilon\eta-24\epsilon^{2}-2\zeta^{2}.
\end{equation}
and for the exact case
\begin{eqnarray}
\alpha_{s}=\frac{(32\xi(\sqrt{\Psi}-2)-192\xi^{2})}{\sqrt{\Psi}(1-\sqrt{\Psi})}\bigg(-\frac{8\xi}{\sqrt{\Psi}}
+\frac{4\xi\bigg((-1+\sqrt{\Psi})^{2}-(1+6\xi)^{2}\bigg)}{(1+6\xi)\Psi}\bigg)\hspace{2cm}\nonumber\\
-\frac{24\bigg(2\xi(\sqrt{\Psi}-2)-12\xi^{2}\bigg)^{2}}{\Psi(1-\sqrt{\Psi})^{2}}-\frac{81M^{2}}{2\pi^{2}\kappa^{4}\xi^{2}}(-1+\sqrt{\Psi})
^{2}(\frac{\kappa^{2}}{6(\sqrt{\Psi}+6\xi)})^{3}\frac{\Psi^{2}(1+6\xi)^{5}}{(-2-6\xi+\sqrt{\Psi})^{4}}.\hspace{1cm}
\end{eqnarray}
The second order of the running of spectral index is defined as
\begin{eqnarray}
\alpha_{s}=16\epsilon\eta-24\epsilon^{2}-2\zeta^{2}-(\frac{428}{3}-48c)\epsilon^{2}\eta-(8c-\frac{92}{3})\epsilon\eta^{2}\hspace{2cm}\nonumber\\
-5(c-\frac{11}{3})\epsilon\zeta^{2}-\frac{1}{2}(7-c)\eta\zeta^{2}-\frac{8}{3}(18c-44)\epsilon^{3}.\hspace{0.5cm}
\end{eqnarray}
again by substitution of the eqs. (22) to (24) and (28) in the above equation, one can obtain
the form of the second order of the running spectral index. The last important inflationary
parameter is the tensor-to-scalar ratio that it is defined as
\begin{equation}
r=16\frac{A_{T}^{2}}{A_{s}^{2}}
\end{equation}
where
\begin{equation}
A_{s}^{2}=\frac{1}{2}\bigg(\frac{H^{2}}{\dot{\varphi}}\big)^{2},\,\,\,\,A_{T}^{2}=\frac{8}{m_{pl}^{2}}\bigg(\frac{H}{2\pi^{2}}\big)^{2}.
\end{equation}
and for this case
\begin{equation}
r=\frac{64}{m_{pl}^{2}}\bigg(\frac{4\xi\bigg[(-1+\sqrt{\Psi})^{2}-(1+6\xi)^{2}\bigg]^{2}}{\kappa^{2}\Psi(-1+\sqrt{\Psi})(1+6\xi)^{3}}\bigg).
\end{equation}
Note that in the above equations, $\Psi$ takes the following form
\begin{equation}
\Psi\equiv1+2\beta^{2}+12\xi\beta^{2}+8\xi N+48\xi^{2}
N+\beta^{4}+12\xi\beta^{4}+36\xi^{2}\beta^{4}.
\end{equation}
\begin{center}
$\bullet\textit{\textbf{Case of $m\gg\beta$}}$
\end{center}
By setting the condition of this case on the equation (27), the form of $m^{2}$ is found as
\begin{equation}
m^{2}=-\frac{1}{(1+6\xi)}\pm\frac{1}{(1+6\xi)}\sqrt{1+8\xi
N+48\xi^{2}N}.
\end{equation}
The inflationary parameters in this limited case is the same with
the previous case. But, the form of $\Psi$ is as
\begin{equation}
\Psi\equiv1+8\xi N+48\xi^{2}N.
\end{equation}
\begin{center}
$\bullet\textit{\textbf{Case of $m\ll\beta$}}$
\end{center}
By adopting the equation (27) with the condition, we have
\begin{equation}
m^{2}=\frac{8\xi N+2\beta^{2}+(1+6\xi)\beta^{4}}{2}.
\end{equation}
The first order of spectral index (29) in this case is driven as
\begin{equation}
n_{s}=1-\frac{24\xi(2-\Sigma)}{
\Sigma(2+(1+6\xi)\Sigma)}-\frac{32\xi}{(2+(1+6\xi)\Sigma)}+\frac{8\xi(1+6\xi)(\Sigma^{2}-4)}{(2+(1+6\xi)\Sigma)^{2}}
\end{equation}
and the first order of the running of spectral index (32) takes the
following form as
\begin{eqnarray}
\alpha_{s}=\frac{64\xi(2-\Sigma)}{\Sigma(2+(1+6\xi)\Sigma)}\bigg(-\frac{16\xi}{(2+(1+6\xi)\Sigma)}+\frac{4\xi(1+6\xi)(\Sigma^{2}-4)}{(2+(1+6\xi)\Sigma)^{2}}\bigg)
\hspace{4cm}\nonumber\\
-24\bigg(\frac{4\xi(2-\Sigma)}{\Sigma(2+(1+6\xi)\Sigma)}\bigg)^{2}
-\frac{162M^{2}\Sigma^{2}}{16\pi^{2}\kappa^{4}\xi^{2}}(\frac{\kappa^{2}}{3(2+\Sigma)})^{3}(\frac{2+(1+6\xi)\Sigma}{\Sigma-2})^{4}.\hspace{0.5cm}
\end{eqnarray}
By applying the eqs. (22) to (24) and (41) in the eqs. (31) and (34), we attain the form
of the second orders of spectral index and the running spectral index, respectively. Also, the
tensor-to-scalar ratio in this case follows as
\begin{equation}
r=\frac{64}{m_{pl}^{2}}\bigg(\frac{4\xi(\Sigma^{2}-4)^{2}}{2\kappa^{2}\Sigma(2+(1+6\xi)\Sigma)^{2}}\bigg)
\end{equation}
where
\begin{equation}
\Sigma\equiv8\xi N+2\beta^{2}+(1+6\xi)\beta^{4}.
\end{equation}
\subsubsection{Power-Law Potential}
The power-law potential
\begin{equation}
V(\varphi)=\lambda\varphi^{n}
\end{equation}
is one of the most significant potentials in the inflationary literatures. In this potential, $\lambda$ is
the dimensionless self-interacting constant and also we are interested for $n\geq4$. It should be
kept in the mind that potential for $n>6$  gives rise to power-law inflation. The field equations can be found as
\begin{equation}
H^{2}\approx\frac{\lambda\kappa^{2}\varphi^{n}}{3(1+\kappa^{2}\xi\varphi^{2})},\,\,
3H\dot{\varphi}\approx\frac{\bigg[(4-n)\kappa^{2}\xi\lambda\varphi^{n+1}-n\lambda\varphi^{n-1}\bigg]}{1+\kappa^{2}\xi\varphi^{2}({1}+6\xi)}
\end{equation}
and by substitution of these equations into the equation (13), the
slow-roll parameters for this potential are demonstrated as
\begin{equation}
\epsilon=\bigg[\frac{\kappa^{4}\xi^{2}\varphi^{2}(n^{2}-6n+8)+2\kappa^{2}\xi(n^{2}-3n)+n^{2}\varphi^{-2}}{2\kappa^{2}({1}+\kappa^{2}\xi\varphi^{2}({1}+6\xi))}\bigg],
\end{equation}
\begin{eqnarray}
\eta=\bigg[\frac{\xi(10n^{2}-3n^{3})}{({1}+\kappa^{2}\xi\varphi^{2}({1}+6\xi))}+\frac{\kappa^{2}\xi^{2}\varphi^{2}(-3n^{3}+14n^{2}-8n-16)}
{({1}+\kappa^{2}\xi\varphi^{2}({1}+6\xi))}+\frac{\kappa^{4}\xi^{3}\varphi^{4}(-n^{3}+6n^{2}-8n)}
{({1}+\kappa^{2}\xi\varphi^{2}({1}+6\xi))}\nonumber\\
+\frac{\varphi^{-2}(2n^{2}-n^{3})}{\kappa^{2}({1}+\kappa^{2}\xi\varphi^{2}({1}+6\xi))}+\frac{\kappa^{2}\xi^{2}\varphi^{2}(1+6\xi)(6n^{2}-12n)}
{({1}+\kappa^{2}\xi\varphi^{2}({1}+6\xi))^{2}}+\frac{\kappa^{6}\xi^{4}\varphi^{6}(1+6\xi)(2n^{2}-12n+16)}
{({1}+\kappa^{2}\xi\varphi^{2}({1}+6\xi))^{2}}\nonumber\\
+\frac{\kappa^{4}\xi^{3}\varphi^{4}(1+6\xi)(6n^{2}-24n+16)}
{({1}+\kappa^{2}\xi\varphi^{2}({1}+6\xi))^{2}}+\frac{2n^{2}\xi(1+6\xi)}{({1}+\kappa^{2}\xi\varphi^{2}({1}+6\xi))^{2}}\bigg]\frac{1}
{(\kappa^{2}\xi\varphi^{2}(2-n)-n)}\hspace{0.5cm}
\end{eqnarray}
and
\begin{equation}
\zeta=\frac{9n}{2\pi}\sqrt{\lambda}\varphi^{(\frac{n}{2}+1)}(\frac{\kappa^{2}}{3(1+\kappa^{2}\xi\varphi^{2})})^{3/2}\bigg[\frac{{1}
+\kappa^{2}\xi\varphi^{2}({1}+6\xi)}{(4-n)\kappa^{2}\xi\varphi^{2}-n}\bigg]^{2}.
\end{equation}
For obtaining the value of $\lambda$, we engage the COBE
normalization as
\begin{equation}
\bigg(\frac{\lambda}{\xi^{2}}\bigg)^{\frac{1}{2}}\approx\frac{1}{2}\bigg(\frac{\delta\rho}{\rho}\bigg)_{hor}=\bigg(\frac{\delta
T}{T}\bigg)_{rms}=1.1\times10^{-5}
\end{equation}
which for the value of $\xi$ gives
$\lambda={\cal{O}}(10^{-12})$. By putting $\epsilon=1$ in the equation (48), for the end of inflation we find $\beta$ as
\begin{equation}
\beta^{2}=\frac{(\frac{1}{\xi}+3n-n^{2})}{(n^{2}-6n+8-\frac{2(1+6\xi)}{\xi})}\pm\frac{\sqrt{13n^{2}+\frac{6n}{\xi}+\frac{1}{\xi^{2}}}}{(n^{2}-6n+8-\frac{2(1+6\xi)}
{\xi})}.
\end{equation}
The slow-roll parameters can be rewritten at the time of HC as
follows
\begin{equation}
\epsilon=\xi\bigg[\frac{m^{4}(n^{2}-6n+8)+2m^{2}(n^{2}-3n)+n^{2}}{2m^{2}({1}+m^{2}({1}+6\xi))}\bigg],
\end{equation}
\begin{eqnarray}
\eta=\bigg[\frac{\xi(10n^{2}-3n^{3})}{({1}+m^{2}({1}+6\xi))}+\frac{\xi m^{2}(-3n^{3}+14n^{2}-8n-16)}
{({1}+m^{2}({1}+6\xi))}+\frac{\xi m^{4}(-n^{3}+6n^{2}-8n)}
{({1}+m^{2}({1}+6\xi))}\nonumber\\
+\frac{\varphi_{HC}^{-2}(2n^{2}-n^{3})}{\kappa^{2}({1}+m^{2}({1}+6\xi))}+\frac{\xi m^{2}(1+6\xi)(6n^{2}-12n)}
{({1}+m^{2}({1}+6\xi))^{2}}+\frac{\xi m^{6}(1+6\xi)(2n^{2}-12n+16)}
{({1}+m^{2}({1}+6\xi))^{2}}\nonumber\\
+\frac{\xi m^{4}(1+6\xi)(6n^{2}-24n+16)}
{({1}+m^{2}({1}+6\xi))^{2}}+\frac{2n^{2}\xi(1+6\xi)}{({1}+m^{2}({1}+6\xi))^{2}}\bigg]\frac{1}
{(m^{2}(2-n)-n)}\hspace{0.5cm}
\end{eqnarray}
and
\begin{equation}
\zeta=\frac{9n}{2\pi}\sqrt{\lambda}\varphi_{HC}^{(\frac{n}{2}+1)}(\frac{\kappa^{2}}{3(1+m^{2})})^{3/2}\bigg[\frac{{1}+m^{2}({1}+6\xi)}{(4-n)m^{2}-n}\bigg]^{2}.
\end{equation}
For this potential, the equation (25) for $N$ is translated as
\begin{equation}
N=\frac{1}{8\xi}\bigg\{\ln\bigg[\frac{(1+m^{2})((4-n)\beta^{2}-n)}{(1+\beta^{2})((4-n)m^{2}-n)}\bigg]+(1+6\xi)
\bigg[\ln(\frac{\beta^{2}+1}{m^{2}+1})+\frac{n}{4-n}\ln\bigg(\frac{(4-n)\beta^{2}-n}{(4-n)m^{2}-n}\bigg)\bigg]\bigg\}
\end{equation}
and by expanding and keeping the first terms, we have
\begin{equation}
N\approx\frac{m^{2}-\beta^{2}}{2\xi n}.
\end{equation}
Now, we consider three confined cases such as the previous potential.
\begin{center}
$\bullet\textit{\textbf{Exact Case}}$
\end{center}
By rewriting the equation (57), we find
\begin{equation}
m^{2}=\beta^{2}+2\xi nN
\end{equation}
and by putting the above equation and the eqs. (53) to (55) in the eqs. (29), (31), (32) and
(34), one can reach to the first order of spectral index, the second order of spectral index,
the first order of the running spectral index and the second order of the running spectral index,
respectively. Also, by using the eqs. (35) and (36), the tensor-to-scalar ratio in this limited
case takes the following form
\begin{equation}
r=\frac{64}{m_{pl}^{2}}\bigg[\frac{(1+\beta^{2}+2\xi n
N)^{2}((\beta^{2}+2\xi n
N)(4-n)-n)^{2}\xi}{\kappa^{2}(\beta^{2}+2\xi n
N)(1+(\beta^{2}+2\xi n N)(1+6\xi))^{2}}\bigg].
\end{equation}
\begin{center}
$\bullet\textit{\textbf{Case of $m\gg\beta$}}$
\end{center}
The form of $m^{2}$ in this case, by applying the condition is
given as
\begin{equation}
m^{2}=2\xi nN.
\end{equation}
The form of all inflationary parameters in this case are similar to the last case if we omit $\beta^{2}$.
In the limited case of $m\ll\beta$, there is no inflationary analysis because there is no inflationary solution.
\subsection{Einstein Frame}
By applying the conformal transformation (2), the conformal factor
is defined as $1+\kappa^{2}\xi\varphi^{2}$. Therefore, the form of
action in this frame is found as follows
\begin{equation}
S_{E}=\int
d^{4}x\sqrt{-\hat{g}}\bigg(\frac{\hat{R}}{2\kappa^{2}}-\frac{1}{2}F^{2}(\varphi)\hat{g}^{\mu\nu}\partial_{\mu}\varphi\partial_{\nu}\varphi
+\hat{V}(\hat{\varphi})\bigg).
\end{equation}
We notice that all quantities in the Einstein frame are labeled by
hat. The scalar field in this frame is redefined as
\begin{equation}
F^{2}(\varphi)\equiv\bigg(\frac{d\hat{\varphi}}{d\varphi}\bigg)^{2}\equiv\frac{1+\kappa^{2}\xi\varphi^{2}({1}+6\xi)}{(1+\kappa^{2}\xi\varphi^{2})^{2}}
\end{equation}
and also we deal with an effective potential
\begin{equation}
\hat{V}(\hat{\varphi})\equiv\frac{V(\varphi)}{(1+\kappa^{2}\xi\varphi^{2})^{2}}.
\end{equation}
By considering FRW metric, the line element is found as
\begin{equation}
d\hat{s}^{2}=d\hat{t}^{2}-\hat{a}^{2}(\hat{t})d\vec{x}^{2}
\end{equation}
where $\hat{a}=\sqrt{\Omega}a$ and $d\hat{t}=\sqrt{\Omega}dt$.
Also, the field equations in this frame take the familiar form of GR
\begin{equation}
\hat{H}^{2}=\frac{\kappa^{2}}{3}\bigg[\Big(\frac{d\hat{\varphi}}{d\hat{t}}\Big)^{2}+\hat{V}(\hat{\varphi})\bigg],
\,\,\,\,\,\,\,\,\,\,\,\frac{d^{2}\hat{\varphi}}{d\hat{t}^{2}}+3\hat{H}\frac{d\hat{\varphi}}{d\hat{t}}+\frac{d\hat{V}}{d\hat{\varphi}}=0.
\end{equation}
Moreover, the quantities in two frames are connected each other by
the following equations
\begin{equation}
\hat{H}\equiv\frac{1}{\hat{a}}\frac{d\hat{a}}{d\hat{t}}=\frac{1}{\sqrt{\Omega}}
\Big(H+\frac{1}{2}\frac{\dot{\Omega}}{\Omega}\Big)
\end{equation}
and
\begin{equation}
\frac{d\hat{\varphi}}{d\hat{t}}\equiv\Big(\frac{d\hat{\varphi}}{d
\varphi}\Big)\Big(\frac{dt}{d\hat{t}}\Big)\dot{\varphi}=\frac{\sqrt{1+\kappa^{2}\xi\varphi^{2}(1+6\xi)}}{\Omega^{\frac{3}{2}}}\dot{\varphi}
\end{equation}
Since in the Einstein frame the effective potential plays an important role, we consider the
slow-roll approximation that is built on potential. The slow-roll conditions are defined as
\begin{equation}
\dot{\hat{\varphi}}^{2}\ll\hat{V},\quad\quad\ddot{\hat{\varphi}}\ll3\hat{H}\dot{\hat{\varphi}}
\end{equation}
and also we can define the slow-roll parameters as
\begin{equation}
\hat{\epsilon}\equiv\frac{1}{2\kappa^{2}}\Big(\frac{\hat{V}^{\prime}(\hat{\varphi})}{\hat{V}(\hat{\varphi})}\Big)^{2},\quad
\hat{\eta}\equiv\frac{1}{\kappa^{2}}\Big(\frac{{\hat{V}}''(\hat{\varphi})}{\hat{V}(\hat{\varphi})}\Big),\quad
\hat{\zeta}\equiv\frac{1}{\kappa^{2}}
\Big(\frac{\hat{V}^{\prime}(\hat{\varphi}){\hat{V}}'''(\hat{\varphi})}{\hat{V}^{2}({\hat\varphi})}\Big)^{\frac{1}{2}}
\end{equation}
where prime implies to the differentiation with respect to the
redefined scalar field $\hat{\varphi}$. We notice that in the Einstein frame,
inflation ends when $\hat{\xi}=1$ or $\hat{\eta}=1$. Such as the Jordan frame, we
consider two usual large field potentials.
\subsubsection{\textbf{Chaotic Inflation Potential}}
The form of effective potential is expressed as
\begin{equation}
\hat{V}(\hat{\varphi})=\frac{M^{2}\varphi^{2}}{2({1}+\kappa^{2}\xi\varphi^{2})^{2}}
\end{equation}
and also we find the slow-roll parameters as
\begin{equation}
\hat{\epsilon}=\frac{2({1}-\kappa^{2}\xi\varphi^{2})^{2}}{\kappa^{2}\varphi^2({1}+\kappa^{2}\xi\varphi^{2}({1}+6\xi))},
\end{equation}
\begin{eqnarray}
\hat{\eta}=-\frac{{16}\xi}{({1}+\kappa^{2}\xi\varphi^{2}({1}+6\xi))}+\frac{{2}({1}+\kappa^{2}\xi\varphi^{2})^2}{\kappa^{2}\varphi^{2}({1}
+\kappa^{2}\xi\varphi^{2}({1}+6\xi))}-\frac{{2}\xi({1}+{6}\xi)({1}
-\kappa^{4}\xi^{2}\varphi^{4})}{({1}+\kappa^{2}\xi\varphi^{2}({1}+6\xi))^2},
\end{eqnarray}
and
\begin{eqnarray}
\hat{\zeta}^{2}=-\frac{64\xi(1-\kappa^{2}\xi\varphi^{2})^{2}}{\kappa^{2}\varphi^{2}({1}+\kappa^{2}\xi^{2}({1}+{6}\xi))^{2}}
-\frac{16\xi(1+6{\xi})(1-4\kappa^{2}\xi\varphi^{2})}{\kappa^{2}\varphi^{2}({1}+\kappa^{2}\xi^{2}({1}+{6}\xi))^{3}}\hspace{3cm}\nonumber\\
+\frac{16\kappa^{2}\xi^{3}\varphi^{2}(1+6\xi)(1-4\kappa^{2}\xi\varphi^{2})}{({1}+\kappa^{2}\xi^{2}({1}+{6}\xi))^{3}}+\frac{16\xi^{2}(1+6\xi)^{2}(1+\kappa^{8}\xi^{4}
\varphi^{8})}{({1}+\kappa^{2}\xi\varphi^{2}({1}+{6}\xi))^{4}}
-\frac{32\kappa^{4}\xi^{4}\varphi^{4}(1+6\xi)^{2}}{({1}+\kappa^{2}\xi\varphi^{2}({1}+{6}\xi))^{4}}.
\end{eqnarray}
To have a graceful exit from inflationary phase, we need to
$\hat{\epsilon}=1$. Hence, by combination of this condition and
eqs. (19) and (71), we can find $\beta$ as the following form
\begin{equation}
\beta^{2}=-\frac{1}{2}\pm\frac{1}{2}\sqrt{\frac{1+12\xi}{1+4\xi}}.
\end{equation}
If we rewrite the slow roll parameters for the beginning of inflation, we have
\begin{equation}
\hat{\epsilon}=\frac{2\xi({1}-m^{2})^{2}}{m^2({1}+m^{2}({1}+6\xi))}
\end{equation}
\begin{eqnarray}
\hat{\eta}=-\frac{{16}\xi}{({1}+m^{2}({1}+6\xi))}
-\frac{{2}\xi({1}+{6}\xi)({1}-m^{4})}{({1}+m^{2}({1}+6\xi))^2}
+\frac{{2}\xi({1}+m^{2})^2}{m^{2}({1}+m^{2}({1}+6\xi))}
\end{eqnarray}
and
\begin{eqnarray}
\hat{\zeta}^{2}=-\frac{64\xi^{2}(1-m^{2})^{2}}{m^{2}({1}+m^{2}(1+6\xi))^{2}}
-\frac{16\xi^{2}(1+6\xi)(1-4m^{2})}{m^{2}({1}+m^{2}(1+6\xi))^{3}}\hspace{3cm}\nonumber\\
+\frac{16m^{2}\xi^{2}(1+6\xi)(1-4m^{2})}{({1}+m^{2}(1+6\xi))^{3}}+\frac{16\xi^{2}(1+6\xi)^{2}(1+m^{8})}{({1}+m^{2}(1+6\xi))^{4}}
-\frac{32\xi^{2}m^{4}(1+6\xi)^{2}}{({1}+m^{2}(1+6\xi))^{4}}.\hspace{0.5cm}
\end{eqnarray}
Also, the number of e-folds can be introduced as
\begin{equation}
\hat{N}=\int^{\hat{t}_{f}}_{\hat{t}_{i}}\hat{H}d\hat{t}=\int^{\hat{\varphi}_{e}}_{\hat{\varphi}_{i}}\frac{\hat{H}}{\dot{\hat{\varphi}}}d\hat{\varphi}.
\end{equation}
also the number of e-folds in two frames are connected each other by
\begin{equation}
\hat{N}=N+\int^{\varphi_{f}}_{\varphi{i}}\frac{\kappa^{2}\xi\varphi}{(1+\kappa^{2}\xi\varphi^{2})}d\varphi.
\end{equation}
If we apply the equation (78) for this potential, we arrive to
\begin{eqnarray}
\hat{N}=\frac{1}{8\xi}\ln\bigg({\frac{(\beta^{2}-1)(m^{2}+1)}{(\beta^{2}+1)(m^{2}-1)}}\bigg)
+\frac{(1+6\xi)}{8\xi}\ln\bigg({\frac{\beta^{4}-1}{m^{4}-1}}\bigg)+
\frac{1}{2}\ln(\frac{\beta^{2}+1}{m^{2}+1})
\end{eqnarray}
and by using the expansion and keeping the first order terms
\begin{equation}
\hat{N}\approx\frac{(1-2\xi)}{4\xi}(m^{2}-\beta^{2})+\frac{(1+6\xi)}{8\xi}(m^{4}-\beta^{4}).
\end{equation}
The strategy for this frame is like the Jordan frame, so we consider three limited cases again.
\begin{center}
$\bullet\textit{\textbf{Exact case}}$
\end{center}
For obtaining the form of $m^{2}$ in this case, the above equation is rewritten as
\begin{eqnarray}
m^{2}=-\frac{(1-2\xi)}{(1+6\xi)}
\pm\frac{1}{(1+6\xi)}\hspace{6cm}\nonumber\\
\times\sqrt{1+4\xi^{2}-4\xi+2\beta^{2}+8\xi\beta^{2}+8\xi
\hat{N}+12\xi\beta^{4}+36\xi^{2}\beta^{4}+48\xi^{2}\hat{N}+\beta^{4}-24\xi^{2}\beta^{2}}.\hspace{0.1cm}
\end{eqnarray}
Now, we can obtain the inflationary parameters for this confined case. The first order of spectral
index in this frame defines as
\begin{equation}
\hat{n_{s}}=1-6\hat{\epsilon}+2\hat{\eta}
\end{equation}
where in this case can be translated as
\begin{eqnarray}
\hat{n_{s}}=1-\frac{12\xi(2+4\xi-\sqrt{\sigma})^{2}}{(-1+2\xi+\sqrt{\sigma})(2\xi+\sqrt{\sigma})(1+6\xi)}-\frac{32\xi}{(2\xi+\sqrt{\sigma})}\hspace{4cm}\nonumber\\
-\frac{4\xi\bigg((1+6\xi)^{2}-(-1+2\xi+\sqrt{\sigma})^{2}\bigg)}{(2\xi+\sqrt{\sigma})^{2}(1+6\xi)}
+\frac{4\xi(8\xi+\sqrt{\sigma})^{2}}{(-1+2\xi+\sqrt{\sigma})(2\xi+\sqrt{\sigma})(1+6\xi)}.\hspace{0.5cm}
\end{eqnarray}
The second order of the spectral index can introduce as
\begin{equation}
\hat{n_{s}}=1-6\hat{\epsilon}+2\hat{\eta}+\frac{1}{3}(44-18c)\hat{\epsilon}^{2}
+(4c-14)\hat{\epsilon}\hat{\eta}+\frac{2}{3}\hat{\eta}^{2}+\frac{1}{6}(13-3c)\hat{\zeta}^{2}.
\end{equation}
By applying the eqs. (75) to (77) and (82), one reach to the form of the second order of spectral
index. Also, the first order of the running of spectral index in this frame can be defined as follows
\begin{equation}
\hat{\alpha_{s}}=16\hat{\epsilon}\hat{\eta}-24\hat{\epsilon}^{2}-2\hat{\zeta}^{2}
\end{equation}
where in this confined case
\begin{eqnarray}
\hat{\alpha_{s}}=\frac{32\xi(2+4\xi-\sqrt{\sigma})^{2}}{(-1+2\xi+\sqrt{\sigma})(2\xi+\sqrt{\sigma})(1+6\xi)}\bigg[-\frac{16\xi}{2\xi+\sqrt{\sigma}}
-\frac{2\xi\bigg((1+6\xi)^{2}-(-1+2\xi+\sqrt{\sigma})^{2}\bigg)}{(2\xi+\sqrt{\sigma})^{2}(1+6\xi)}\nonumber\\
+\frac{2\xi(8\xi+\sqrt{\sigma})^{2}}{(-1+2\xi+\sqrt{\sigma})(2\xi+\sqrt{\sigma})(1+6\xi)}\bigg]
-\frac{96\xi^{2}(2+4\xi-\sqrt{\sigma})^{4}}{(-1+2\xi+\sqrt{\sigma})^{2}(2\xi+\sqrt{\sigma})^{2}(1+6\xi)^{2}}\hspace{1cm}\nonumber\\
+\frac{128\xi^{2}(2+4\xi-\sqrt{\sigma})^{2}}{(-1+2\xi+\sqrt{\sigma})(2\xi+\sqrt{\sigma})^{2}(1+6\xi)}
+\frac{32\xi^{2}(5-2\xi-4\sqrt{\sigma})(1+6\xi)}{(-1+2\xi+\sqrt{\sigma})(2\xi+\sqrt{\sigma})^{3}}\hspace{1cm}\nonumber\\
-\frac{32\xi^{2}(-1+2\xi+\sqrt{\sigma})(5-2\xi-4\sqrt{\sigma})}{(1+6\xi)(2\xi+\sqrt{\sigma})^{3}}
-\frac{32\xi^{2}\bigg((1+6\xi)^{4}+(-1+2\xi+\sqrt{\sigma})^{4}\bigg)}{(1+6\xi)^{2}(2\xi+\sqrt{\sigma})^{4}}\nonumber\\
+\frac{64\xi^{2}(-1+2\xi+\sqrt{\sigma})^{2}}{(2\xi+\sqrt{\sigma})^{4}}.\hspace{0.5cm}
\end{eqnarray}
The second order of the running spectral index in the Einstein frame can be expressed as
\begin{eqnarray}
\hat{\alpha}_{s}=14\hat{\epsilon}\hat{\eta}-12\hat{\epsilon}^{2}-2\hat{\zeta}^{2}-2(-18c+\frac{151}{3})\hat{\epsilon}^{2}\hat{\eta}-2
(-\frac{44}{3}+4c)\hat{\epsilon}\hat{\eta}^{2}\hspace{2cm}\nonumber\\
-5(c-\frac{11}{3})\hat{\epsilon}\hat{\zeta}^{2}-\frac{1}{2}(7-c)\hat{\eta}\hat{\zeta}^{2}-\frac{4}{3}(44-18c)\hat{\epsilon}^{3}\hspace{0.5cm}
\end{eqnarray}
by using the slow-roll parameters in this case, one can obtain the form of the second order of
the running spectral index. The form of the tensor-to-scalar ratio defines as
\begin{equation}
\hat{r}=16\frac{\hat{A}_{T}^{2}}{\hat{A}_{s}^{2}}
\end{equation}
where
\begin{equation}
\hat{A}_{s}^{2}=\frac{1}{2}\bigg(\frac{\hat{H}^{2}}{\dot{\hat{\varphi}}}\big)^{2},\,\,\,\,\hat{A}_{T}^{2}=\frac{8}{m_{pl}^{2}}\bigg
(\frac{\hat{H}}{2\pi^{2}}\big)^{2}.
\end{equation}
and for this case
\begin{equation}
\hat{r}=\frac{256(2\xi+\sqrt{\sigma})(1+6\xi)}{\kappa^{2}m_{pl}^{2}(-1+2\xi+\sqrt{\sigma})}\bigg(\frac{(-2-4\xi+\sqrt{\sigma})}
{(1+6\xi)(2\xi+\sqrt{\sigma})+2\xi(-2-4\xi+\sqrt{\sigma})}\bigg)^{2}.
\end{equation}
Note that in the above equations, $\sigma$ takes the following form
\begin{equation}
\sigma\equiv1+4\xi^{2}-4\xi+2\beta^{2}+8\xi\beta^{2}+8\xi \hat{N}+12\xi\beta^{4}+36\xi^{2}
\beta^{4}+48\xi^{2} \hat{N}+\beta^{4}-24\xi^{2}\beta^{2}.
\end{equation}
\begin{center}
$\bullet\emph{\textbf{Case of $m\gg\beta$}}$
\end{center}
By setting the condition of this case on the equation (81), the form of $m^{2}$ is found as
\begin{equation}
m^{2}=-\frac{(1-2\xi)}{(1+6\xi)}\pm\frac{1}{(1+6\xi)}\sqrt{1+4\xi^{2}-4\xi+8\xi
\hat{N}+48\xi^{2}\hat{N}}.
\end{equation}
All inflationary parameters in this case are the same with the previous case with different form
of $\sigma$ as
\begin{equation}
\sigma\equiv1+4\xi^{2}-4\xi+8\xi \hat{N}+48\xi^{2} \hat{N}.
\end{equation}
\begin{center}
$\bullet\emph{\textbf{Case of $m\ll\beta$}}$
\end{center}
By adopting the equation (81) with the condition, we have
\begin{equation}
m^{2}=\frac{8\xi \hat{N}+(2-4\xi)\beta^{2}+(1+6\xi)\beta^{4}}{(2-4\xi)}.
\end{equation}
The expression for the first order of spectral index (83) follows as
\begin{eqnarray}
\hat{n}_{s}=1-\frac{12\xi(2-4\xi-\Theta)^{2}}{\Theta(2-4\xi+(1+6\xi)\Theta)}-\frac{32\xi(2-4\xi)}{(2-4\xi+(1+6\xi)\Theta)}\hspace{4cm}\nonumber\\
+\frac{4\xi(2-4\xi-\Theta)^{2}}{\Theta(2-4\xi+(1+6\xi)\Theta)}-\frac{4\xi(1+6\xi)((2-4\xi)^{2}-\Theta^{2})}{(2-4\xi+(1+6\xi)\Theta)^{2}}.
\end{eqnarray}
By engaging the eqs. (75) to (77) and (95) in the eqs. (85), (86) and (88), one can obtain the
second order of spectral index and the first and second orders of the running spectral index,
respectively. For the tensor-to-scalar ratio
\begin{equation}
\hat{r}=\frac{256\xi((2-4\xi)+\Theta(1+6\xi))}{\kappa^{2}m_{pl}^{2}\Theta}\bigg(\frac{(\Theta-(2-4\xi))}{((2-4\xi)+\Theta(1+6\xi)+
2\xi(\Theta-(2-4\xi)))}\bigg)^{2}
\end{equation}
where
\begin{equation}
\Theta\equiv\beta^{2}(2-4\xi)+8\xi \hat{N}+(1+6\xi)\beta^{4}.
\end{equation}
\subsubsection{\textbf{Power-Law Potential}}
The definition (63) for this potential can be written as
\begin{equation}
\hat{V}(\hat{\varphi})=\frac{\lambda\varphi^{n}}{({1}+\kappa^{2}\xi\varphi^{2})^{2}}
\end{equation}
and by combination of the above equation and the equation (69), the slow-roll parameters take
the following forms
\begin{equation}
\hat{\epsilon}=\frac{(n+(n-4)\kappa^{2}\xi\varphi^{2})^{2}}{2\kappa^{2}\varphi^{2}({1}+\kappa^{2}\xi\varphi^{2}({1}+6\xi))}
\end{equation}
\begin{eqnarray}
\hat{\eta}=\frac{n(n-1)}{\kappa^{2}\varphi^{2}({1}+\kappa^{2}\xi\varphi^{2}({1}+6\xi))}+\frac{\kappa^{2}\xi^{2}\varphi^{2}(n^{2}-7n+12)}
{({1}+\kappa^{2}\xi\varphi^{2}({1}+6\xi))
}+\frac{2\xi(n^{2}-4n-2)}{({1}+\kappa^{2}\xi\varphi^{2}({1}+6\xi))}\hspace{2cm}\nonumber\\
+\frac{\kappa^{4}\xi^{3}\varphi^{4}(1+6\xi)(4-n)}{({1}+\kappa^{2}\xi\varphi^{2}({1}+6\xi))^{2}}+\frac{2\kappa^{2}\xi^{2}\varphi^{2}(1+6\xi)(2-n)}
{({1}+\kappa^{2}\xi\varphi^{2}({1}+6\xi))^{2}}-\frac{\xi
n(1+6\xi)}{({1}+\kappa^{2}\xi\varphi^{2}({1}+6\xi))^{2}}.
\end{eqnarray}
By setting $\hat{\epsilon}=1$ to the end of inflation, the form of $\beta$ follow as
\begin{equation}
\beta^{2}=\frac{\bigg(-(1-n(n-4)\xi)\pm\sqrt{1+8\xi
n+12\xi^{2}n^{2}}\bigg)}{(2(1+6\xi)-\xi(n-4)^{2})}.
\end{equation}
For the start of inflation, the slow-roll parameters are rewritten as following terms
\begin{equation}
\hat{\epsilon}=\frac{\xi(n+(n-4)m^{2})^{2}}{2m^{2}({1}+m^{2}({1}+6\xi))}
\end{equation}
\begin{eqnarray}
\hat{\eta}=\frac{n(n-1)\xi}{m^{2}({1}+m^{2}({1}+6\xi))}+\frac{\xi
m^{2}(n^{2}-7n+12)}{({1}+m^{2}({1}+6\xi))
}+\frac{2\xi(n^{2}-4n-2)}{({1}+m^{2}({1}+6\xi))}\hspace{4cm}\nonumber\\
+\frac{\xi
m^{4}(1+6\xi)(4-n)}{({1}+m^{2}({1}+6\xi))^{2}}+\frac{2\xi
m^{2}(1+6\xi)(2-n)}
{({1}+\kappa^{2}\xi\varphi^{2}({1}+6\xi))^{2}}-\frac{\xi
n(1+6\xi)}{({1}+m^{2}({1}+6\xi))^{2}}.\hspace{0.5cm}
\end{eqnarray}
Also, the number of e-folds can be driven as
\begin{eqnarray}
\hat{N}=\frac{1}{8\xi}\bigg\{\ln\bigg[\frac{(1+m^{2})((4-n)\beta^{2}-n)}{(1+\beta^{2})((4-n)m^{2}-n)}\bigg]\hspace{4cm}\nonumber\\
+(1+6\xi)\bigg[\ln(\frac{\beta^{2}+1}{m^{2}+1})+\frac{n}{4-n}\ln\bigg(\frac{(4-n)\beta^{2}-n}{(4-n)m^{2}-n}\bigg)\bigg]\bigg\}+\frac{1}{2}
\ln(\frac{\beta^{2}+1}{m^{2}+1})\hspace{0.5cm}
\end{eqnarray}
and by expanding the above equation
\begin{equation}
\hat{N}\approx\frac{(m^{2}-\beta^{2})(1-n\xi)}{2n\xi}.
\end{equation}
\begin{center}
$\bullet\textit{\textbf{Exact Case}}$
\end{center}
In this limited case, the equation (106) can be rewritten as
\begin{equation}
m^{2}=\beta^{2}+\frac{2n\xi\hat{N}}{1-n\xi}.
\end{equation}
and by setting the above equation and the eqs. (103) and (104) in the eqs. (83), (85), (86) and
(88), we can achieve to the first order of spectral index, the second order of spectral index,
the first order of the running spectral index and the second order of the running spectral index,
respectively. Also, by using the eqs. (89) and (90), the tensor-to-scalar ratio follows as
\begin{equation}
\hat{r}=\frac{64\xi}{\kappa^{2}m_{pl}^{2}\Lambda((1-n\xi)+\Lambda(1+6\xi)}\bigg(\frac{\Lambda(4-n)-n(1-n\xi)}{1+\frac{\xi(\Lambda(4-n)-n(1-n\xi))}{((1-n\xi
)+\Lambda(1+6\xi))}}\bigg)^{2}
\end{equation}
where
\begin{equation}
\Lambda\equiv(1-n\xi)\beta^{2}+2n\xi \hat{N}.
\end{equation}
\begin{center}
$\bullet\textit{\textbf{Case of $m\gg\beta$}}$
\end{center}
We can express the form of $m^{2}$ as
\begin{equation}
m^{2}=\frac{2n\xi\hat{N}}{1-n\xi}.
\end{equation}
also all inflationary parameters in this case are the same with the previous case if we neglect $\beta^{2}$ in the equations. In the case of $m\ll\beta$ such as the Jordan frame, there is no inflationary
solutions for this potential.

\section{The Analysis of Obtained Results in Two Frames}
In this section, we analysis our results in two frames for both potentials by comparing with the latest observational datasets. The Planck satellite reveals the spectral index as $n_{s} = 0.9652 \pm 0.0047$, the running spectral index as $\frac{dn_{s}}{dlnk} =-0.022 \pm 0.010$ and the tensor-to-scalar ratio as $r < 0.099$. Tables 1 and 2 show the obtained results for chaotic inflation potential in Jordan and Einstein frames, respectively. By comparing two tables, we can see that in the exact case, the first order of spectral index for both frames have big difference and also the numbers in the Einstein frame have much better consistency with observational data. For the second order of spectral index, the discrepancy between two frames is preserved and again the Einstein frame has better condition than the Jordan frame. About the running spectral index, tables show that the obtained results for the first and second orders are different in two frames and the values in the Einstein frame is more close to observations. Also, the tables present the values for tensor-to-scalar ratio in two frames. In the next limited case, the analysis is almost similar to the previous case. For the third specific case, the story is different than other cases. The tables reveal that the values of the first and second orders of spectral index are the same in both frames, nearly. In other words, the Jordan and Einstein frames are almost equivalent. But, there are some differences between two frames for running spectral index. Moreover, it seems that the Einstein frame has better agreement than the Jordan frame in this case. In addition to the above consequences, we can see that for all three cases the values for $\xi=\frac{1}{12}$ are closer to the observations rather than $\xi=\frac{1}{6}$.\\
Another large field potential for inflation is power-law potential. To analysis the issue, the calculation are done in case of $n = 4$. The tables 3 and 4 present the values for inflationary parameters in both frames for different cases. For the first order of spectral index in the exact case, two frames have insignificant difference and the Jordan frame has better agreement with observational datasets rather than the Einstein frame. Also, they have maintained the difference for the second order of spectral index with different values. Likewise, the tables reveal that running spectral index is negative for this potential and again the Jordan frame has better consistency with observational data than the Einstein frame. For the next case, values and situation between frames are almost the same with the exact case. Such as the last potential, the values for $\xi=\frac{1}{12}$ is much better than $\xi=\frac{1}{6}$ .
\section{Discussion and Outlook}
In the previous sections, the obtained results for two large field potentials in non-minimal case compared for Jordan and Einstein frames. Now, we are going to discuss about these consequences and its relation with the issue of frames as the main purpose of this paper. As we know the chaotic and power-law potentials as large field potentials have engaged in wide range of inflationary scenarios. In principle, the form of potential is $V(\varphi)=\frac{1}{2}M^{2}\varphi^{2}+\lambda\varphi^{4}+V_{0}$ where two first terms are very usual among people. Therefore, we can also compare the results for two potentials from this viewpoint. Based on the results, the chaotic potential as a large field potential, not only has some values in case of $m\gg\beta$ also it has some values in case of $m\ll\beta$. In other words, inflaton shows a derivation from its nature so that it has twofold behaviour. But, we cannot see this derivation for the power-law potential. Hence, it seems that the power law potential can reveal the nature of large field potentials better than chaotic potential. Now, we can look to these consequences from two frames approach. According to former works that have done about the issue of frames, Jordan and Einstein frames which are connected each other by conformal transformations are physically equivalent and at the presences of quantum corrections of inflaton, we might find some discrepancies between two frames. The results in the previous section for two potentials present that the physically equivalent for two frames in chaotic potential can be found only in the case of $m\ll\beta$ and the mentioned derivation plays an important role for this potential. But, for power-law potential, the difference between two frames is not very significant and we can claim both frames are nearly equivalent. Despite the above results, if we focus on disagreement between two frames in two potentials, some interesting results can be found about the issue of frames. The values for inflationary parameters in tables claim that for the chaotic potential, the Einstein frame has better agreement with observations than the Jordan frame and this situation is different for the power-law potential so that the Jordan frame is physical frame respect to the Einstein frame. It seems that the behaviour of inflaton decides about the type of physical frame. The above analysis dissolve that the dual behaviour of inflaton in the chaotic potential is mysterious. Therefore, finding its nature can be very useful to raise of our knowledge from the nature of inflaton. In order to clarify this issue, a few points might be considered, briefly. It is very well to keep in our mind that the mass term in the form of chaotic potential might has strange influences in inflationary scenarios particularly at the presence of quantum corrections of scalar field. Also, we know that inflaton in the chaotic potential behaves like a harmonic oscillator, so it might be possible to assume a dual manner for inflaton. Another important case to point out is that we usually avoid from considering some vital factors in our inflationary situations in order to smooth the model. As example, we can remark to the role of temperature in the early universe era or some features for perfect fluid like viscosity and some feasible corrections from the string theory that can be added to the model as the perturbations of our inflationary scenario.\\
We know that the issue of frames is an unresolved problem in GR and there are some attempts to clarify the issue by means of some gravitational scenarios. But, people usually tend to avoid from involving with this ambiguity in the papers. The present paper attempt to describe some
hidden aspects of the issue of frames in order to illuminate the future landscape. Because of this, the author would like to express some important points that might be effective to have better perception about the problem. Despite the above discussion about the issue of frames,
some people do not believe to this issue they think the Jordan and Einstein frames are always equivalent physically because of the covariance principle in GR. But, some theoretical evidences show that the existence of difference between two frames is not out of mind. An example, we
know that in some quantum situations the equivalence principle of GR will be broken because in quantum level we need to Quantum Gravity (QG) theory to describe the universe. Therefore, breaking other principles of GR such as covariance principle might be happened in quantum
level. In principle, to present a better judgement about physical frame between Jordan and Einstein frames, it is better to assume this issue in the context of quantum gravitational theory or at least in the presence of some quantum corrections of the model. It is obvious that QG as
usual quantum gravitational theory is not available for us, so talking about the frames problem is possible only by considering some quantum corrections.
\section{Summary}
The issue of frames as an old open question in GR considered in the present work and we attempted to clarify this problem by non-minimal inflationary models. According to the strategy, firstly we presented non-minimal inflationary analysis and then introduced Jordan and Einstein frames. We continued the analysis for two usual large field potentials with different limited cases in two frames, separately. In the next step, we compared the calculated inflationary parameters for two potentials in both frames with observations in order to clear the
issue of frames. The obtained results showed that there are some derivations for the chaotic inflation potential in case of $m\ll\beta$, however we can not see the derivation for another large field potential. Also, we realized that by accepting the difference between two frames, for the
chaotic inflation potential the Einstein frame is the physical frame and for power-law potential the Jordan frame is the physical frame. Moreover, we presented a quantitative discussion about the obtained results and some important points as outlook. As we told the quantum corrections can have significant role in the issue of frames, so it might be interesting for the future works. Because of this reason, we propose some quantum considerations to clarify the issue of frames by non-minimal inflation. First suggestion is using the deformation of phase space of classical gravitational theory by non-commutatively of mini-superspaces. We know that there are many physical theories that are considered from non-commutative approach. Among those, we are interested about the effects of non-commutativity upon cosmological scenarios by the formulation
of a version of non-commutative cosmology in which a deformation of mini-superspace is required. In fact, by using deformation mechanism, we map from the classical phase space to quantum phase space of theory and it is expected that some new features can be available than
the classical case. Another proposal to consider quantum corrections for the issue of frames is applying the Finsler geometry as a general form of Riemannian geometry in our cosmological scenarios. We know that the Riemannian geometry as a base of GR have a great role in our
cosmological scenarios. The standard cosmological model has achieved many successes to explain the universe and some observations can support them, but there are some cosmological phenomenas that GR is unable to illuminate them such as Dark Energy (DE) problem. Because of this, some people prefer to modify the GR by producing some changes in gravity sector of the Einstein field equation as modified theories of gravity. Another group of people are interested to make some changes in the matter sector of the Einstein field equation by considering some
new components for the universe. After Discovering DE, both approaches are conventional in cosmological papers. There is a different viewpoint with two mentioned approaches which is based on some fundamental changes in GR. According to this, it seems that making some
changes in background geometry can open new windows to us in order to clarify the mysteries of the universe. The candidate for this important work is Finsler geometry. There are many attempts to apply this geometry for various aspects of cosmology, however they are restricted to use the effective form of this geometry only by using Finsler metric instead FRW metric in Riemannian background. In fact, the transition from Riemannian to Finsler geometry basically is a fundamental work and it is very difficult to perform, so people are forced to apply this
geometry effectively.\\\\
{\bf Acknowledgement}\\\\
I would like to thank Professor Kourosh Nozari beacause of his useful hints.\\

\newpage
\begin{table}
\begin{center}
\caption{Obtained result for Chaotic inflation potential in Jordan
frame}\vspace{0.5cm}
\begin{tabular}{|||c|c|c|c|c|c|c|||}
  \hline
  $$ & $\xi$ & $n_{s}$ & $n_{s}$& $\alpha_{s}$ & $\alpha_{s}$ & $r$ \\
  $$ & $$ & $(1st\,\,Order)$ & $(2nd\,\,Order)$ & $(1st\,\,Order)$ & $(2nd\,\,Order)$ & $$\\
  \hline
  $$& $\frac{1}{6}$ & $0.90434$ & $0.92091$ & $0.027957$ & $0.025557$ & $0.0017893$ \\
  $Exact\,\,Case$& $$ & $$ & $$ & $$ & $$ & $$ \\
  \cline{2-7}
  $$ & $\frac{1}{12}$ & $0.91319$ & $0.92441$ & $0.018145$ & $0.017049$ & $0.0030348$ \\
  $$& $$ & $$ & $$ & $$ & $$ & $$ \\
  \hline
  $$& $\frac{1}{6}$ & $0.90393$ & $0.92061$ & $0.028115$ & $0.025699$ & $0.0018051$ \\
  $m\gg\beta$ & $$ & $$ & $$ & $$ & $$ & $$ \\
  \cline{2-7}
  $$ &  $\frac{1}{12}$ & $0.91272$ & $0.92402$ & $0.018262$ & $0.017159$ & $0.0030665$\\
  $$& $$ & $$ & $$ & $$ & $$ & $$ \\
  \hline
  $$& $\frac{1}{6}$ & $0.99081$ & $0.99115$ & $0.00067568$ & $0.00066268$ & $0.0000063619$ \\
  $m\ll\beta$& $$ & $$ & $$ & $$ & $$ & $$ \\
  \cline{2-7}
  $$ &  $\frac{1}{12}$ & $0.98672$ & $0.98733$ & $0.0011578$ & $0.0011303$ & $0.000043325$\\
  $$& $$ & $$ & $$ & $$ & $$ & $$ \\
  \hline
\end{tabular}
\end{center}
\end{table}

\begin{table}
\begin{center}
\caption{Obtained result for Chaotic inflation potential in Einstein
frame}\vspace{0.5cm}
\begin{tabular}{|||c|c|c|c|c|c|c|||}
  \hline
  $$ & $\xi$ & $n_{s}$ &$n_{s}$& $\alpha_{s}$ & $\alpha_{s}$ & $r$ \\
  $$ & $$ & $(1st\,\,Order)$ & $(2nd\,\,Order)$ & $(1st\,\,Order)$ & $(2nd\,\,Order)$ & $$\\
  \hline
  $$& $\frac{1}{6}$ & $0.97511$ & $0.99272$ & $0.015719$ & $0.010016$ & $0.017044$ \\
  $Exact\,\,Case$& $$ & $$ & $$ & $$ & $$ & $$ \\
  \cline{2-7}
  $$ & $\frac{1}{12}$ & $0.97418$ & $0.98627$ & $0.010485$ & $0.0076732$ & $0.019758$ \\
  $$& $$ & $$ & $$ & $$ & $$ & $$ \\
  \hline
  $$& $\frac{1}{6}$ & $0.97501$ & $0.99268$ & $0.01578$ & $0.010052$ & $0.017108$ \\
  $m\gg\beta$& $$ & $$ & $$ & $$ & $$ & $$ \\
  \cline{2-7}
  $$ &  $\frac{1}{12}$ & $0.97404$ & $0.98619$ & $0.010538$ & $0.0077091$ & $0.019867$\\
  $$& $$ & $$ & $$ & $$ & $$ & $$ \\
  \hline
  $$& $\frac{1}{6}$ & $0.99973$ & $0.99991$ & $0.00018133$ & $0.00017203$ & $0.00017188$ \\
  $m\ll\beta$& $$ & $$ & $$ & $$ & $$ & $$ \\
  \cline{2-7}
  $$ &  $\frac{1}{12}$ & $0.998896$ & $0.99938$ & $0.0004857$ & $0.0075991$ & $0.00077726$\\
  $$& $$ & $$ & $$ & $$ & $$ & $$ \\
  \hline
\end{tabular}
\end{center}
\end{table}

\begin{table}
\begin{center}
\caption{Obtained result for Power-low potential in Jordan frame
for $n=4$ }\vspace{0.5cm}
\begin{tabular}{|||c|c|c|c|c|c|c|||}
  \hline
  $$ & $\xi$ & $n_{s}$ & $n_{s}$& $\alpha_{s}$ & $\alpha_{s}$ & $r$ \\
  $$ & $$ & $(1st\,\,Order)$ & $(2nd\,\,Order)$ & $(1st\,\,Order)$ & $(2nd\,\,Order)$ & $$\\
  \hline
  $$& $\frac{1}{6}$ & $0.97467$ & $0.97352$ & $-0.0051174$ & $-0.0051965$ & $0.021297$ \\
  $Exact\,\,Case$& $$ & $$ & $$ & $$ & $$ & $$ \\
  \cline{2-7}
  $$ & $\frac{1}{12}$ & $0.96584$ & $0.9652993$ & $-0.00096384$ & $-0.000990898$ & $0.037903$ \\
  $$& $$ & $$ & $$ & $$ & $$ & $$ \\
  \hline
  $$& $\frac{1}{6}$ & $0.97444$ & $0.97331$ & $-0.00495596$ & $-0.0050335$ & $0.021496$ \\
  $m\gg\beta$& $$ & $$ & $$ & $$ & $$ & $$ \\
  \cline{2-7}
  $$ &  $\frac{1}{12}$ & $0.96542$ & $0.96486$ & $-0.00097458$ & $-0.0010024$ & $0.038366$\\
  $$& $$ & $$ & $$ & $$ & $$ & $$ \\
  \hline
  $$& $\frac{1}{6}$ & $---$ & $---$ & $---$ & $---$ & $---$ \\
  $m\ll\beta$& $$ & $$ & $$ & $$ & $$ & $$ \\
  \cline{2-7}
  $$& $\frac{1}{12}$ & $---$ & $---$ & $---$ & $---$ & $---$\\
  $$& $$ & $$ & $$ & $$ & $$ & $$ \\
  \hline
\end{tabular}
\end{center}
\end{table}

\begin{table}
\begin{center}
\caption{Obtained result for Power-low potential in Einstein frame
for $n=4$}\vspace{0.5cm}
\begin{tabular}{|||c|c|c|c|c|c|c|||}
  \hline
  $$ & $\xi$ & $n_{s}$ & $n_{s}$& $\alpha_{s}$ & $\alpha_{s}$ & $r$ \\
  $$ & $$ & $(1st\,\,Order)$ & $(2nd\,\,Order)$ & $(1st\,\,Order)$ & $(2nd\,\,Order)$ & $$\\
  \hline
  $$& $\frac{1}{6}$ & $0.99443$ & $0.99376$ & $-0.0036337$ & $-0.0036291$ & $0.000058721$ \\
  $Exact\,\,Case$& $$ & $$ & $$ & $$ & $$ & $$ \\
  \cline{2-7}
  $$ & $\frac{1}{12}$ & $0.98505$ & $0.98396$ & $-0.0059873$ & $-0.0059695$ & $0.00061825$ \\
  $$& $$ & $$ & $$ & $$ & $$ & $$ \\
  \hline
  $$& $\frac{1}{6}$ & $0.99442$ & $0.99374$ & $-0.0036427$ & $-0.003638$ & 0.000059018\\
  $m\gg\beta$& $$ & $$ & $$ & $$ & $$ & $$ \\
  \cline{2-7}
  $$ &  $\frac{1}{12}$ & $0.98494$ & $0.98385$ & $-0.0060248$ & $-0.0060068$ & $0.00062675$\\
  $$& $$ & $$ & $$ & $$ & $$ & $$ \\
  \hline
  $$& $\frac{1}{6}$ & $---$ & $---$ & $---$ & $---$ &$---$\\
  $m\ll\beta$& $$ & $$ & $$ & $$ & $$ & $$ \\
  \cline{2-7}
  $$ &  $\frac{1}{12}$ & $---$ & $---$ & $---$ & $---$ & $---$\\
  $$& $$ & $$ & $$ & $$ & $$ & $$ \\
  \hline
\end{tabular}
\end{center}
\end{table}
\end{document}